# Characterizing Pedophile Conversations on the Internet using Online Grooming

Aditi Gupta, Ponnurangam Kumaraguru, Ashish Sureka

Indraprastha Institute of Information Technology
New Delhi - 110078, INDIA
{aditig, pk, ashish}@iiitd.ac.in
http://precog.iiitd.edu.in/

*Abstract*—Cyber-crime targeting children such as online pedophile activity are a major and a growing concern to the society. A deep understanding of predatory chat conversations on the Internet has implications in designing effective solutions to automatically identify malicious conversations from regular conversations. We believe that a deeper understanding of the pedophile conversation can result in more sophisticated and robust surveillance systems than majority of the current systems relying only on shallow processing such as simple word-counting or key-word spotting.

In this paper, we study pedophile conversations from the perspective of online grooming theory and perform a series of linguistic-based empirical analysis on several pedophile chat conversations to gain useful insights and patterns. We manually annotated 75 pedophile chat conversations with six stages of online grooming and test several hypothesis on it. The results of our experiments reveal that *relationship forming* is the most dominant online grooming stage in contrast to the *sexual stage*. We use a widely used word-counting program (LIWC) to create psycho-linguistic profiles for each of the six online grooming stages to discover interesting textual patterns useful to improve our understanding of the online pedophile phenomenon. Furthermore, we present empirical results that throw light on various aspects of a pedophile conversation such as probability of state transitions from one stage to another, distribution of a pedophile chat conversation across various online grooming stages and correlations between pre-defined word categories and online grooming stages.

## I. INTRODUCTION

Protection of children on cyber space is an extremely critical problem faced by our society across geographical and cultural boundaries. As, more and more children in their teens have started using the Internet, there has been an alarming increase in cases of child abuse through the Web. One of the reports published by National Centre for Missing and Exploited Children (NCMEC), reported that 1 in 7 kids is solicited for sex online; 1 in 33 kids receives aggressive online solicitation to meet in person; 1 in 3 kids receives unsolicited sexual content online [10]. The Internet provides criminals easy and anonymous access to children. Criminals exploit the fact that children in adolescent years are curious and inquisitive to obtain more information about topics like sex, drugs, etc. Some ways by which criminals are targeting children are: exploit children (sexual abuse); cyber-bully children (emotional abuse); corrupt children (violence abuse); distract children (social abuse) [15]. Problems related to child abuse need to be dealt with extreme sensitivity and using a focussed framework built specifically to deter, detect and defend crimes against children. By doing this, we believe that the Internet can be used for good and better by children.

Among various problems faced by children online, one of the fastest growing problem is Pedophile. According to Diagnostic and Statistical Manual of Mental Disorders, *pedophillia is characterized by either intense sexually arousing fantasies, urges, or behaviors involving sexual activity with a prepubescent child (typically age 13 or younger)*. The individuals categorized as pedophiles by above definition must be at least 16 years old and at least 5 years older than the child. [1]

Pedophile exploit children through various means on the Internet like websites, blogs, online-forums and chat-rooms. Penna et al. outlines the research need to build tools and techniques for automatic detection and analysis of pedophile activities on the Internet [9]. Our ongoing research aims to build the framework for a real time automated tool that can flag an ongoing conversation on the Internet as Pedophile conversation. The framework will utilize text analytics and machine learning models which can be deployed in a chat room monitoring software or parental control plug-ins in instant messaging or chat clients. We aim to advance state-of-the-art in non-topical text classification models. In this paper, we discuss the first step towards building the system – characterize / profile online grooming theory from Psychology, identify the indicators that characterize the various stages of online grooming. The aim of profiling these stages is to build a system (classifier) to identify the presence of grooming in a chat, and hence serve as an indicator to the chat being between a pedophile and a minor. Though that is the ultimate goal, in this paper, we just present our initial work towards that goal.

In order to understand a pedophile's behavioral and linguistic patterns of conversation with a minor, we use the *theory of online grooming* from psychology as our foundation. The contribution of this paper is an empirical analysis on pedophile chat conversations to understand and gain novel insights into the online grooming practices. The following summarize some

---
[1]http://allpsych.com/disorders/paraphilias/pedophilia.html

of the key results obtained in this paper:
- Our results reveal that in a pedophile conversation *relationship forming* is the most prominent stage out of the six stages of online grooming. This is contrary to belief, as the pedophile conversations are expected to majorly contain sexual content.
- We observe that pedophiles do not wait for the end of grooming process to approach a child to meet in person, as was suggested in psychology literature. On the contrary, we show that *conclusion stage* should be considered as a basic and recurrent stage like the other five stages of online grooming.
- We investigate the linguistic styles and create psycholinguistic profiles for each of the six online grooming stages using logistic regression analysis. Developing the profiles for each stage can be used to design an automatic text-based classifier to detect stages of online grooming.

Developing text analytics solutions for chat conversations over the Internet is a challenging problem. This is so due to the informal nature and manner of language, grammar and vocabulary used in a chat. Hence, data cleaning and pre-processing becomes absolutely essential. Also, due to the informal language and grammar usage in a text chat, application of standard models and techniques of text analytics cannot be used. Some non-technical challenges include obtaining dataset containing conversations by pedophiles, because it is a sensitive social issue special care need to be taken in anonymizing the data. Also, heterogeneity in the format in which logging of chat is done by various Internet chat rooms and messengers pose a challenge to an automated system.

The rest of the paper is organized in the following manner: In section II, we give the related work and background about work done in the fields of computer science, psychology and criminology related to pedophile activities on the Internet and the theory on online grooming from psychology. In Section III, we discuss the approaches we took in analyzing the chats, and descriptive statistics about the dataset that we analyzed. In Section IV, we discuss results which help us in characterizing online grooming in pedophile chat conversations. In Section V, we look at the implications of the results and we discuss some of the limitations and future directions that we plan to take in this research.

## II. BACKGROUND

### A. Related Work

Chat room and instant message monitoring in context to the problem of predicting child-sex related solicitation conversation has received some interest in recent times. But it is still a relatively unexplored and uncharted area given the magnitude and seriousness of the problem. On some aspects work has been done in Computer Science, e.g. analyzing chat conversations and websites. Researchers used simple text mining and statistical techniques specifically directed towards chat data analysis for detecting illegitimate conversation related to pedophile activity [6], [7], [13]. Traditional solutions have limitations, they do not do deep linguistic analysis as a result of which more research is needed to improve existing models and techniques. Researchers have also used social network analysis of child exploitation websites, Frank et al. extracted structure and features of four online child exploitation networks [4].

Pendar et al. investigated the feasibility of categorizing the text chat of the perpetrator (the sexual predator) and the victim by applying SVM and kNN based classification models. Pendar et al. takes a bag-of-word approach and extract n-gram word sequences (unigrams, bigrams and trigrams for their experiments) from the Perverted Justice corpus for the purpose of building a predictive model. The study presents encouraging results with the f-measure of 0.943 on test dataset for the problem of discriminating child and victim chat messages [13]. However, our interest is on investigating the applicability of sophisticated semantic-based features which can facilitate in building a more advanced model of the communication and intent complementing existing approaches and make them more robust towards false alarms. Thus, the major difference between the study done by Pendar et al. and this paper is that, in this work, we investigate the presence of certain semantic features leading to a different direction than the study done by Pendar et al.

Kontostathis et al. proposed an approach based on theoretical foundations drawn from Olson et al.'s communication theory of child sexual predators [12]. Kontostathis et al. propose salient features and lexical clues (for building a classification model to flag such entrapping online chat conversations) based on adapting the communication process (stages and intentions) in physical-world to the online-world. Kontostathis et al. analyze the text conversation and tag the presence of communication phases such as gaining access, deceptive trust development, grooming, isolation, and approach using hand-created lexicon and rules. The basis of the approach proposed by Kontostathis et al. is in luring communication theory which is realized by creating linguistic clues consisting of dictionary of pre-defined terms and phrases and hand-crafted rules [6], [7], [8]. The main difference between the work done by Kontostathis et al. and this paper is that we investigate the presence of different stages and linguistic cues to online grooming which has a different theoretical base than the work of Kontostathis et al.

O'Connell created a typology for online child sex related activities [11]. Gray et al. showed that pedophiles have an association between children and sex [5]. D'Ovidio et al. conducted content analysis of adult-child websites and concluded that they are criminogenic in that they contain a myriad of communication tools (e.g. chat rooms, instant messengers, and message boards) to promote and facilitate pedophile activities on the Internet [14]. The report by Choo on online child grooming, discusses in depth the concept on online grooming; ways in which predators exploit technology to groom children online; impact of these on policing, policymaking and legislation; and suggest ways by which we can respond to this issue [2].

While the theory of online grooming is a well studied concept in psychology to understand online pedophile activity,

the application of online grooming theory for automatically detecting or classifying text-based online pedophile chat conversations is unexplored. To the best of our knowledge, this is the first academic study on performing empirical analysis on pedophile chat conversations using the elements of online grooming theory. A close investigation to learn relationship between various dimensions of word categories (such as rate of emotional and sexual words) and online grooming stages (such as friendship, relationship and risk assessment) is a novel contribution of this paper. We provide answers to research questions such as, which online grooming stage is most dominant (in terms of percentage of messages or time spent) and what is the probability of transitions between different stages. Viewing a chat conversation in terms of a state transition diagram and learning the probability of transitions from one state to another from actual dataset has consequences in designing an automatic text-based pedophile chat conversation classifier which is our ultimate research goal.

*B. Online Grooming: Basic Premise of Proposed Solution and Theoretical Foundations*

Internet provides pedophiles easy and anonymous access to children. Janeiro noted that inexperience and young age of a child is exploited online for grooming them for offline meeting [3]. We take the *theory of online grooming* from psychology as our foundation to study a pedophile's behavioral and linguistic patterns of chat conversation with a child online. Rachel O'Connell categorized online text conversations containing online grooming by a predator, into the following six stages [11]:

- *Friendship forming stage*: The friendship forming stage comprises of conversation in which the pedophile tries to get introduced to the child. The pedophile and victim exchange their *name, location, age, etc*. Generally it follows up pedophile inquiring about *other online accounts information* about the child and *requests pictures* from him in order to confirm that the person on the other side is indeed a child.
- *Relationship forming stage*: In this stage, the pedophile talks about *family and school life* of the child. He tries to know more about the *interest and hobbies* of the child so that he can exploit them by deceptively making the child believe that they are in a relationship.
- *Risk assessment stage*: The pedophile in this stage tries to gauge the level of threat and danger that he is in by talking to a child. He *ensures if the child is alone* or that *any body else is not reading their conversation*. He often asks the child to *delete their chat logs* or confronts the child directly about him being a police officer.
- *Exclusivity stage*: In this stage, the pedophile tries to gain the *trust* of the child completely. He asserts that they share a *special bond*. Often the concept of *love* and *care* are introduced by pedophile in this stage. O'Connell observes that when a pedophile questions a child about how much he trusts him then a child mostly responds to this by admitting that he trusts the adult implicitly.
- *Sexual stage*: In this stage, it often begins with question like *are you a virgin?* or *do you touch yourself?* Some pedophiles talk in great depth about sexual activities with the children to make them accustomed to the language and content. They do this to groom the child for the actual physical interaction.
- *Conclusion stage*: In this stage, the pedophile approaches a child for meeting him in person. He discusses about where they can meet and what activities they would engage when they meet. He asks a child *If he can come to child's home* or something like what would you like to do when we meet and often suggests like *I would like to feel you when I come*.

O'Connell also asserted that these stages of online grooming may or may not be sequential. The frequency, order and extent of occurrence of these stages in a pedophile chat may vary from chat to chat.

### III. METHODOLOGY

*A. Data: Collection and Preprocessing*

Lack of actual and real-world dataset (consisting of chat conversations) for research and evaluation purposes is cited as one of the major bottlenecks (due to natural implications related to privacy) in studying the pedophile problem in greater detail. To the best of our knowledge, the only source that is currently publicly available is the Perverted Justice. [2] Perverted Justice (PJ) is a volunteer center which is part of a not-for-profit foundation focusing efforts aimed at countering crime related to sexual predators against children in particular and pedophiles. Perverted Justice performs sting operations wherein a volunteer poses as a child and engages in a conversation with a suspect on the Internet. The online chat conversation where a suspect tries to solicit a minor is recorded and used as an evidence to convict the perpetrator. Perverted Justice works with law enforcement agencies and their efforts have resulted in 505 convictions as of February 18, 2010 (since June 2004). The count was 288 in Aug 2008, had an increase of 217 from Aug 2008 to Feb 2010. We manually collected the chats from webpages of PJ website. The chat transcripts of all the 502 conversations that we used for analysis are available on PJ's website. [3] The study and experiments performed by Pendar et al. and Kontostathis et al. are on the PJ dataset. Eventhough the dataset is a conversation between a suspect and an undercover agent (pseudo-victim rather than an actual victim), it still offers a starting point for building computational models for detecting and monitoring such conversations. In future, we plan to acquire more realistic dataset (if possible as it poses several non-technical issues) and share it with the broader research community. Table I gives the descriptive statistics for the PJ dataset.

This dataset from perverted justice is a challenging dataset in itself. The chats are copied from various sources together at one place, also the completeness of the chats cannot be verified as original source logs are not made available by the website. The chat logs have comments written by the volunteers in between the chat. Also, the language used in chats is very informal and the vocabulary consists of various slangs, shorthands, emoticons and spelling mistakes. Hence, the data required various steps of preprocessing and cleaning before analysis could be actually performed. Firstly, around 5,100 slangs, short hands and emoticons were replaced in the dataset, Table II gives some examples of these. Secondly, all comments written by the volunteers in between the chats were removed. Thirdly, the chat IDs and timestamp were removed from the chats. So now our dataset consisted of only the lines of actual written words by the victims and predators.

---

[2] http://www.perverted-justice.com/
[3] There were no transcripts for 3 chat conversations at the time when we downloaded the data.

TABLE I
DESCRIPTIVE STATISTICS OF PJ DATASET. VALUES IN BRACKET GIVE THE STANDARD DEVIATION.

| Type of statistics | PJ Dataset | Annotated chats |
|---|---|---|
| Total number of chats | 502 | 75 |
| Total number of lines | 10,46,112 | 47,416 |
| Total number of words | 87,43,941 | 4,02,377 |
| Number of words per line | 8.36 | 8.49 |
| Average lines per chat | 2,082.89 (24,250.49) | 631.3 (1,124.37) |
| Average words per chat | 17,416.16 (24,250.49) | 5,361.96 (9,796.61) |

TABLE II
SAMPLE SLANGS/EMOTICONS IN CHATS.

| Chat-Abbreviations | Full Form |
|---|---|
| a/s/l | Age/Sex/Location |
| BRB | Be Right Back |
| OMG | Oh My God |
| :-) | smile |
| 4ever | forever |
| alrite | Alright |

### B. Annotation

We randomly selected 75 chats out of the 502 chat dataset to get them manually annotated from a professional psychologist in our team for stages of online grooming. These chats comprised 47,416 lines and 4,02,377 words in all. Table III describes in brief the descriptors associated with each stage that were identified while manual annotation and some sample as examples from dataset.

### C. Hypotheses

Since pedophiles are sexual offenders, the conversations they engage in with children online are majorly sexual in nature. Hence detecting *sexual stage* as the central stage, will help in building a system that can take this as an input and flag chat to be pedophile or not. The hypothesis seems quite trivial and obvious, but it would be interesting to observe the degree of support it gets from empirical analysis. As, one would assume that pedophile conversations would mostly contain only sexual content, but O'Connell showed that there are five other stages of grooming present [11]. Though seemingly trivial, it will be significant to empirically obtain the extent and characteristics of the *sexual stage* as most of the solutions developed in the market to detect pedophile activities rely on counting words of sexual nature to flag chats as pedophile. Hence, this hypothesis will help us to assess the effectiveness and strength of such solutions.

**Hypothesis 1:** *Sexual stage of online grooming is the most central stage in pedophile chat conversations.*

O'Connell in her work on online grooming, described the five basic stages of online grooming; the length and order of which varies from chat to chat. She further stated that the pedophile usually ends their conversation with the *conclusion stage* [11]. After the grooming of a child is completed by a pedophile using the five basic stages, then he suggests to the child that they should meet in person.

**Hypothesis 2:** *Pedophile approaches the child to meet in person in the conclusion stage of online grooming.*

Since each stage of online grooming has very clear and distinct descriptors regarding what is discussed in them. It seems intuitive that all stages have characteristic conversation styles, hence we can find linguistic identifiers for each. It would be useful to obtain these indicators as they can be fed into a natural language based machine learning classifier. Building such a classifier to distinguish each stage independently would enable us in detecting presence of online grooming in a text conversation based on the presence of all the stages.

**Hypothesis 3:** *Linguistic predictors can be identified to distinguish each stage of online grooming.*

### D. Analysis

We manually segment a pedophile chat conversation into various stages of online grooming and employ a widely used word counting program (LIWC) [4] to derive correlations between pre-defined word categories and online grooming stages. The correlation between word categories and various stages of online grooming has implications in designing an automatic text-based pedophile chat conversation classifier. Since we divided each chat into the six stages of online grooming we got a working dataset of 450 [5] text scripts in all.

*Detecting transition between stages of online grooming:* We studied the transition in each of the manually annotated chat from one stage of online grooming to the next. We calculated the frequencies for each of the stages. We recorded the above in a 6 X 6 matrix, where rows and columns represented the stages on online grooming. Next for each entry in the table we calculated the corresponding conditional probabilities. An entry $a(i,j)$ in the matrix, denotes conditional probability that the conversation moves to state $Sj$ given the state $Si$. Hence, all entries in the diagonal of the matrix were assigned the values zero. The probabilities were calculated by the following formula:

$$Pr(Sj|Si), Sj = next state, Si = current state$$

---
[4] http://www.liwc.net/
[5] In total 75 chat conversations were annotated for six stages of manual grooming by the psychologist, hence corresponding to each chat we got 6 more text files. Hence 450 = 75 * 6 text scripts are obtained as dataset

TABLE III
DESCRIPTORS FOR STAGES OF ONLINE GROOMING.

| Stages | Friendship forming | Relationship forming | Risk assessment | Exclusivity | Sexual | Conclusion |
|---|---|---|---|---|---|---|
| Descriptor1 | Exchanging email address / picture / web-cam information [early in chat] | Exchanging email address / picture / web-cam information [later in chat] | Checking child's parents are around OR Who all else uses the computer | Feeling of love and exclusiveness expressed | Giving body and figure description | Arranging for a day, date, time and location to meet in person |
| Descriptor2 | Talking about boyfriends / girlfriends [early in chat] | Giving soft compliments, e.g. Sweetie, cutie | Asking the child to delete their chat logs, ensuring no body else had password of child's account | Describing sexual activity and experiences to the child | Becoming boyfriend/ girlfriend of each other | Discussing how to commute to the meeting point |
| Descriptor3 | Getting information about other accounts and online profiles of the child | Taking about a child's hobbies, activities and interests | Checking if the child is fine with seeing an older man/woman | Giving strong compliments, e.g. you are sweetheart | Exchanging pictures of sexual nature or body parts | Ensuring child will come alone to meet |
| Descriptor4 | Asking the age / gender / location / name / personal information / details about family | School / Grade / Homework / cell / phone number | Directly confronting to ensure that child is not a cop/police-agent | Building trust with the child | Giving sexually oriented compliments, e.g. sexy | Deciding on what to do when they meet in person |

*Psycho-Linguistic Profiles for Stages of Online Grooming:* To create psycho-linguistic profiles for each of the six stages, we use the LIWC tool. We ran the LIWC tool on the 450 preprocessed data text files containing only the text of the chat between the predators and victims. LIWC gave us probabilities of frequencies of words in each text across 72 different categories. Since in this study, we focus only on psychological aspects of the conversations between a pedophile and a victim, we consider only the psychological based categories of LIWC. Also, we removed the categories with less than 0.1% value. Finally, we were left with 29 LIWC categories. Next, we calculated the Z-scores for all the LIWC categories to normalize the values belonging to different categories. Using these 29 categories of LIWC as independent variables, we did logistic regression analysis to estimate linguistic predictors for each stage of online grooming. For each stage, we took a dependent variable, with value 1 corresponding to the files belonging to that stage and 0 in the rest. Using this, estimators for each stage were calculated. We calculated the VIF index (VIF stands for variance inflation factor). For the analysis to check for multi-collinearity in the data. [6]. It was observed that there was a high degree of collinearity in the data. This high collinearity in data was occurring due to the fact that many categories of LIWC had overlapping word lists. Some of the word categories were a superset of others in LIWC. Hence, we removed collinearity of data, by not considering the superset category, for the categories with more than 80% overlap. Now, again VIF index was calculated for the 19 categories left. We got all VIF index values to be less than 10 indicating the collinearity was resolved [1].

## IV. RESULTS

Our results from the analysis above partially support Hypothesis 1; completely negate Hypothesis 2; and support Hypothesis 3.

### A. Hypothesis 1

When we plot the pie-chart for the distribution of pedophile conversations into the six stages (based on number of lines of chats in each stage), we get the results as shown in Figure 1. We observe that *friendship forming stage* spans 40% of lines in a pedophile conversation, while *sexual stage* only covers around 24% of the lines. Hence contrary to belief, we observed that *sexual stage* is not the most prominent stage or central stage in a pedophile conversation. Next we calculate the betweenness [7] and closeness centrality [8] of the graph obtained from the 6*6 conditional probability matrix discussed above. The directed graph for the six stages is shown in Figure 2. The betweenness and closeness centrality values obtained for all the six nodes are listed in Table IV. We find various central stages in the graph: Relationship forming and sexual (maximum betweenness centrality); friendship forming and exclusivity (maximum closeness centrality). Thus, sexual stage is not the only central stage according to betweenness centrality. Since sexual stage is the second most dominant stage in a pedophile's conversation and it is one of the central stages, hypothesis 1 is not completely negated, but not supported either.

---

[6]http://en.wikipedia.org/wiki/Multicollinearity: Multicollinearity is a statistical phenomenon in which two or more predictor variables in a multiple regression model are highly correlated. Multicollinearity does not reduce the predictive power or reliability of the model as a whole; it only affects calculations regarding individual predictors

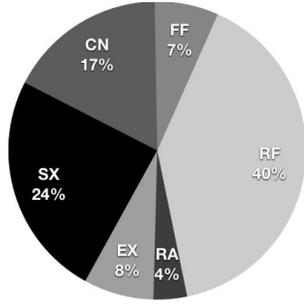

Fig. 1. Distribution of pedophile chat among the six stages of online grooming. The pie-chart clearly shows that the *Relationship forming stage* takes a major portion of pedophile's conversation. Followed by the *Sexual stage* and *Conclusion stage*. FF: Friendship Forming RF: Relationship Forming RA: Risk Assessment EX: Exclusivity SX: Sexual CN: Conclusion.

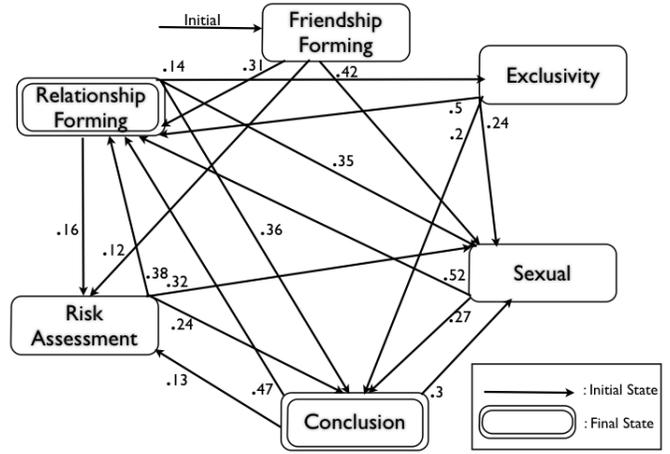

Fig. 2. The flow of conversation between the various stages of online grooming. A node represents each stage in online grooming. An edge from stage $Si$ to $Sj$ represents that stage $Sj$ is followed by stage $Si$ in the pedophile conversation. Weights on the edges represent the conditional probability that the conversation moves to state $Sj$ given state $Si$. Friendship forming stage is the initial stage with which most of pedophile conversations begin. Relationship forming and conclusion stages are the two final stages with which conversations end.

This indicates that tools and solutions developed on simple keyword searching for sexual content to detect pedophile activity in chat rooms may not prove to be very effective. Hence a framework to do more in-depth linguistic analysis based on psychological thinking and behavioral patterns of pedophiles, like online grooming theory is required.

TABLE IV
THE BETWEENNESS AND CLOSENESS CENTRALITY MEASURES FOR THE PEDOPHILE CONVERSATION GRAPH.

| Stage | Betweenness Centrality | Closeness Centrality |
|---|---|---|
| FF | 0 | 13 |
| RF | 1.167 | 11 |
| RA | 0.33 | 12 |
| EX | 0 | 13 |
| SX | 1.167 | 11 |
| CN | 0.33 | 12 |

### B. Hypothesis 2

Figure 1 showed that conclusion stage occupies only 17% of the chat conversations; it is the stage that contains the third largest percentage of words in a pedophile's conversation. Our analysis of transitions in a pedophile conversation from one stage to the other reveals that in all 404 transitions (20% of total transitions) occur from conclusion stage to all other stages. In addition to that, we observe that their was no particular order or position in which the *conclusion stage* occurs. Hence, we assert that contrary to theory, *conclusion stage* should be considered as a basic and recurrent stage like all the other five stages of online grooming. This has implications in designing a real-time system to detect online grooming as, we would check for the presence of *conclusion stage* throughout the predators conversation and not just the end of conversation.

---

[7]Betweenness centrality indicates which nodes have high importance based on location. The node with high betweenness centrality value act as the link between subcomponents of the graph.

[8]Closeness centrality indicates which nodes access most of other nodes. Nodes with high closeness value have shortest paths to maximum other nodes.

### C. Hypothesis 3

By applying logistic regression analysis on the LIWC results for each of the six stages, we found the various categories of words that act as predictors for each stage. Table V presents the outcome of the regression analysis. The *social* and *sexual* words categories demonstrate an interesting pattern. More number of *social* category words indicate the stage of friendship and relationship forming; whereas less number of *social* words indicate either sexual or conclusion categories. *Sexual* category words act as positive predictors for sexual stage and occur in low frequency in relationship forming and conclusion. These results are inline with known knowledge and descriptions of these two stages.

Relationship forming stage has the minimum number of predictors. This is due to the general nature of the conversation that occurs in this stage. This is the stage in which the pedophile spends most part of the conversation discussing about the chid's activities, interests and friends, in an attempt to build a bond with the child. Since this stage has the minimum number of predictors, it would be difficult to identify by a classifier.

TABLE V
PREDICTORS OBTAINED FROM LOGISTIC REGRESSION ANALYSIS. SIGNIF. CODES: 0: *** ; 0.001: ** ; 0.01: * ; (+): INDICATES POSITIVE CORRELATION ; (-): INDICATES NEGATIVE CORRELATION.

| LIWC Categories | Examples | FF | RF | RA | EX | SX | CN |
|---|---|---|---|---|---|---|---|
| Social processes | Mate, talk, they, child | *** (+) | ** (+) | | | **(-) | ***(-) |
| Family | Daughter, husband, aunt | *** (-) | | *** (+) | | *(-) | *(-) |
| Humans | Adult, baby, boy | | | * (+) | | ***(+) | |
| Positive emotion | Love, nice, sweet | | | *** (-) | **(+) | | |
| Negative emotion | Hurt, ugly, nasty | | *(+) | ** (+) | | | ***(-) |
| Insight | Think, know, consider | | | | *(+) | | |
| Causation | Because, effect, hence | | | | *(-) | | |
| Discrepancy | Should, would, could | ***(-) | | | *(+) | *(+) | |
| Inclusive | And, with, include | *** (-) | | * (+) | *** (+) | | |
| Exclusive | But, without, exclude | | | * (+) | | | |
| Perceptual processes | Observing, heard, feeling | *** (+) | | | ** (-) | | **(-) |
| Body | Cheek, hands, spit | | | *** (-) | | ***(+) | |
| Health | Clinic, flu, pill | | *** (+) | | | | *(+) |
| Sexual | Horny, love, incest | | ** (-) | | | ***(+) | **(-) |
| Space | Down, in, thin | ** (+) | | | *** (-) | | **(+) |
| Motion | Arrive, car, go | | | | | | ***(+) |
| Time | End, until, season | | | | | | ***(+) |

Risk assessment stage is categorized by presence of high number of *family* and *negative emotion* words. Greater number of *family* words occur as, in this stage the pedophile focusses to talk about the family and background of the child. In this stage, pedophile tries to ensure from the child that the details of their interaction will remain a secret. This explains the presence of *negative emotion* words, as the pedophile does so by portraying to the child the negative consequences that can occur if he/she (child) revealed any details to anyone.

*Positive emotion* words are present in high numbers in exclusivity stage. Exclusivity is the stage which the pedophile gains complete trust of the child by creating an illusion that he loves and cares for the child, for doing so, he uses a lot of positive emotion words. Certain categories like *motion* and *time* words are very strong indicators for the conclusion stage but do not appear as indicators for any of the other stages. This is so because the pedophile and the child discuss about when and where they will meet and what activities they would engage in. Table V shows the other linguistic profile that we have created for each of the six stages of online grooming.

### D. Observations

Some further inferences we can draw from our analysis are discussed in this section. The piechart in Figure 1 indicates that three stages of friendship, risk assessment and exclusivity all occupy less than 8% each of total pedophile conversation content. Though present at comparatively low levels, these stages are an integral part of online grooming process. From Figure 2, we infer that the stage with which the pedophile initiates his conversation is invariably the friendship stage but the stage with which the conversation end may vary between conclusion and relationship forming stage. The reason for relationship stage to be occurring at the end of the conversation may be explained by the nature of our dataset. Since these chats are between agents pretending to be children to catch the pedophiles, some times the conversations end abruptly as the pedophile gets captured by the agency.

## V. DISCUSSION

In this paper, we studied pedophile conversations from the perspective of online grooming theory from psychology. We performed a series of linguistic-based empirical analysis on several pedophile chat conversations to understand and gain novel insights into the online grooming practices. From our experiments, we concluded that in a pedophile conversation *relationship forming* is the most prominent stage out of the six stages of online grooming. This is contrary to belief, as the pedophile conversations are expected to majorly contain sexual content. It also implies that tools and solutions developed on simple keyword searching for sexual content to detect pedophile activity in chat rooms may not prove to be very effective. Furthermore, we observed that pedophiles do not wait for the end of grooming process to approach a child to meet in person, as was suggested in psychology literature. Hence, while detecting presence of online grooming we should search for presence of *conclusion stage* throughout the predator's conversation and not just the end. We also investigated the linguistic styles and created psycho-linguistic profiles for each of the six online grooming stages using logistic regression analysis. Developing the profiles for each stage can be used to design an automatic text-based classifier to detect stages of online grooming.

### A. Limitations

The dataset used in this work is biased as the agents (pretending as victims) in the dataset are undercover officers; the real dataset for this study would be conversation between a real victim (child) and a pedophile. The characterization of online grooming is currently tested only on the PJ dataset which is known to consist of grooming by pedophiles. Presence or absence of online grooming and its stages in general (non-pedophile) chats still remains an open area of study. But doing that analysis is out of the scope of this paper, and we will continue our research to achieve these in our upcoming work. There will be certain kinds of chat conversations like those between teens or between adults discussing sexual content, that can pose serious challenges to any system built to detect pedophile conversations on the Internet. This is so because the nature, contents and words used in these would be very close to pedophile conversations themselves. Hence, only word counting and statistical techniques may not suffice to distinguish between such adult chats and pedophile chats. We would require a framework to do more in-depth linguistic analysis based on psychological thinking and behavioral patterns of pedophiles, using online grooming theory.

## B. Future work

By building a classifier to detect online grooming, we aim to build an automated system to distinguish a pedophile conversation from a non-pedophile conversation. As part of future work, we plan to build a machine learning classifier which can distinguish between the two kinds of chat conversations. Figure 3 explains the different steps of our project that would lead to achieve that aim.

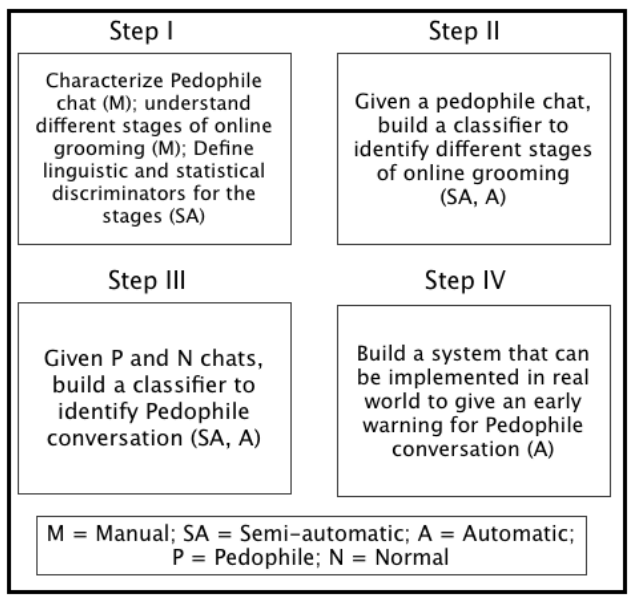

Fig. 3. Architecture of an automated system to distinguish between a pedophile and non-pedophile conversation using online grooming theory. It describes the procedure from characterizing online grooming to building a real-time automated system to distinguish a pedophile conversation from a non-pedophile conversation.

The three contributions of this paper discussed earlier will directly serve as an input to Step II. Once we obtain an automated classifier to distinguish stages of online grooming in a chat conversation, we would extend to a classifier that can distinguish between pedophile conversation and non-pedophile conversation in Step III. The last and ultimate goal as mentioned in Step IV would be to adapt the classifier build a real time early warning system to flag an ongoing chat conversation as pedophile. In the process of achieving above research goals, we also aim to develop publicly available datasets of Internet chat conversations for pedophile and non-pedophile research advancement.

ACKNOWLEDGEMENT

The authors would like to thank all those who helped in annotating the chat transcripts. Special thanks to Dr. Smita Sahay for her inputs on annotation and linking us to the psychology literature. The authors thank all members of PreCog research group at IIIT-Delhi for their support.

REFERENCES

[1] D. Belsley, E. Kuh, and R. Welsch, Regression diagnostics, John Wiley and Sons, 1980.
[2] K. R. Choo, Online child grooming: a literature review on the misuse of social networking sites for grooming children for sexual offenses, Tech. report, Australian Institute of Criminology, 2009.
[3] R. D. Janeiro, Child pornography and sexual exploitation of children online, Tech. report, A contribution of ECPAT International to the World Congress III against Sexual Exploitation of Children and Adolescents, 2008.
[4] R. Frank, B. Westlake, and M. Bouchard,The structure and content of online child exploitation networks, ACM SIGKDD Workshop on Intelligence and Security Informatics (2010), 9.
[5] N. S. Gray, A. S. Brown, M. J. MacCulloch, and J. Smith, An implicit test of the associations between children and sex in pedophiles, Tech. report, Journal of Abnormal Psychology, 2005.
[6] A. Kontostathis, L. Edwards, J. Bayzick, I. Mcghee, A. Leatherman, and K. Moore, Comparison of rule-based to human analysis of chat logs, 2010, Retrieved Aug 16, 2010, http://webpages.ursinus.edu/akontostathis/KontostathisMSMFinal.pdf.
[7] A. Kontostathis, L. Edwards, and A. Leatherman, Chat- coder: Toward the tracking and categorization of internet predators, Text Mining Workshop 2009 held in conjunction with the Ninth SIAM International Conference on Data Mining (2009), 9.
[8] M. W. Berry, J. Kogan, M. W. Berry, J. Kogan, A. Kontostathis, L. Edwards, and A. Leatherman, Text mining and cybercrime, in text mining: Application and theory, Eds., John Wiley and Sons, Ltd., 2009.
[9] L. Penna, A. Clark, and G. Mohay, Challenges of automating the detection of paedophile activity on the internet, First International Workshop on Systematic Approaches to Digital Forensic Engineering on Systematic Approaches to Digital Forensic Engineering (2005).
[10] NCMEC, National center for missing and exploited children, 2008, http://www.missingkids.com/missingkids/servlet/NewsEventServlet?LanguageCountry=en_US&PageId=4303.
[11] R. O'Connell, A typology of child cybersexploitation and online grooming practices, Tech. report, Cyberspace Research Unit, University of Central Lancashier, 2003, Retrieved Aug 12, 2010, http://image.guardian.co.uk/sys-files/Society/documents/2003/07/24/Netpaedoreport.pdf.
[12] L. N. Olson, J. L. Daggs, B. L. Ellevold, and T. K. K. Rogers, Entrapping the innocent: Toward a theory of child sexual predators luring communication, Communication Theory 17(3), 231-251, 2007.
[13] N. Pendar, Toward spotting the pedophile telling victim from predator in text chats, ICSC 07: Proceedings of the International Conference on Semantic Computing (Washington, DC, USA), IEEE Computer Society, 2007, pp. 235241.
[14] R. D'Ovidio, T. Mitman, I. J. El-Burki, and W. Shumar, Adult-child sex advocacy websites as social learning environments: A content analysis, International Journal of Cyber Criminology Vol 3 Issue 1 January (2009).
[15] B. Thuraisingham, Protecting our children in cyberspace, Retrieved Sept 12, 2010, http://www.utdallas.edu/ bxt043000/Motivational-Articles/motivational-articles.html.